          [2001/04/25 1.1 (PWD)]
\documentclass[a4paper,twocolumn]{spaceDebrisC} 
\usepackage{times}
\usepackage[numbers]{natbib}
\usepackage{graphicx}
\usepackage{hyperref}

\title{An integrated debris environment assessment model} 
\author{Akhil Rao}
\affil{Middlebury College Department of Economics, Middlebury VT 05753, USA, Email: akhilr$@$middlebury.edu}
\author{Francesca Letizia}
\affil{IMS Space Consultancy at ESA/ESOC, Robert-Bosch-Str. 5, 64293 Darmstadt, Germany, Email: francesca.letizia$@$esa.int}

\begin{document}


\maketitle

\begin{abstract}
Launch behaviors are a key determinant of the orbital environment. Physical and economic forces such as fragmentations and changing launch costs, or policies like post-mission disposal (PMD) compliance requirements, will alter the relative attractiveness of different orbits and lead operators to adjust their launch behaviors. However, integrating models of adaptive launch behavior with models of the debris environment remains an open challenge. We present a statistical framework for integrating theoretically-grounded models of launch behavior with evolutionary models of the low-Earth orbit (LEO) environment. We implement this framework using data on satellite launches, the orbital environment, launch vehicle prices, sectoral revenues, and government budgets over 2007-2020. The data are combined with a multi-shell and multi-species Particle-in-a-Box (PIB) model of the debris environment and a two-stage budgeting model of commercial, civil government, and defense decisions to allocate new launches across orbital shells. We demonstrate the framework's capabilities in three counterfactual scenarios: unexpected fragmentation events in highly-used regions, a sharp decrease in the cost of accessing lower parts of LEO, and increasing compliance with 25-year PMD guidelines. Substitution across orbits based on their evolving characteristics and the behavior of other operators induces notable changes in the debris environment relative to models without behavioral channels. 
\end{abstract}

\section{Introduction}

Economic and technological trends have led to a rapid expansion in orbit use \cite{envRep2020}. Policies to manage the increase in space traffic may be developed based on projections of the future orbital environment, which in turn depend on projections of future launch behavior. Understanding the ways orbit users respond to the changing state of the orbital environment is key to developing cost-effective policies which are robust to likely behavioral responses. 

We present a framework for integrating behavioral models of orbit use with physical models of orbital environment evolution. The framework is focused on launch decisions, leaving satellite design and disposal decisions as exogenous parameters. Satellite operators in the model respond to the evolving environment in ways consistent with their observed preferences over orbital characteristics and observed or modeled changes in economic and policy parameters. Unlike repeating historical launch patterns, this approach allows analysts to observe how satellite operators will likely substitute over orbits in response to evolving conditions when launching new satellites, and study the influences of global space-related spending and the state of the orbital environment on the number of satellites launched each year. This framework can be used to prospectively assess policy effectiveness, value orbital characteristics from operators' perspectives, and assess environmental sensitivity to behavioral or economic factors.

Prior work integrating economic and debris environment models focused on optimization models of launch behavior \cite{adilov2020model, raoetal2020, rouillon2020economic}. While optimization models facilitate optimal policy design, they can be computationally burdensome to solve, requiring many simplifications which reduce physical realism and obscure important choice margins. The statistical framework applied here extends the approaches in prior works, allowing for more physical realism and more choice margins. Future iterations will also enable statistical inference on satellite operators' valuations of orbital characteristics (e.g. estimates of willingness-to-pay for debris removal, economic damages suffered due to fragmentation events net of substitution across orbits). The number of satellites launched annually by each operator (user) type (commercial, civil government, and defense) and their distribution across orbits is projected using a two-stage budgeting econometric model, e.g. \cite{jorgensen1988budgeting, hausman1995utility}. The econometric model is embedded within a debris environment model which propagates the evolution of the debris environment in response to physical processes and economic or policy choices. 

It is important to emphasize that this framework is not intended to make forecasts of the debris environment. While future iterations may provide useful forecasts, research is still needed to better understand the appropriate model structures and specifications for forecasting. Rather, the framework is intended to address two goals. First, to illustrate a way to construct quantitative debris environment models which integrate behavioral and physical dimensions. Second, to illustrate the kinds of responses and patterns which may result from sudden or gradual changes to the orbital environment or the economics of orbit use.

\section{The integrated model framework}

The framework has three components:
\begin{enumerate}
    \item A model of the debris environment, which takes as inputs the current physical state of the orbital region and the distribution of satellites being launched in that period. This is referred to as the {\bf debris environment model}.
    \item A model of operators' choices to launch satellites to specific shells, which takes as input the current physical state of the orbital region and some economic parameters such as access costs. This is referred to as the {\bf second-stage model}.
    \item A model of the total number of satellites launched by each type of operator, which takes as input a summary of the current physical state of the orbital environment as a whole, the current distribution of revenues across different sectors of the space economy, measures of government space budgets, and a price index derived from the model of choice over orbits. This is referred to as the {\bf first-stage model}.
\end{enumerate}

The first-stage and second-stage models together define our econometric model of operators' demand for orbits. This two-stage econometric approach is common in natural resource modeling to jointly study substitution over resource sites and entry to resource use, e.g. \cite{hausman1995utility}. It is also widely used to study other kinds of consumer choice, e.g. \cite{deaton1980aids, freeman2014measurement}.


\subsection{Data}
\label{data}
We use data on objects in orbit and satellites launches from the DISCOS (Database and Information System Characterising Objects in Space) database \cite{DISCOS}. These data describe the distributions of active satellites and debris in each altitude shell by operator category and object type, and the population of objects launched to each shell in each year. These data also classify each launch by operator category and list the average active lifetime of satellites by operator category. The operator categories and object types are described in section \ref{pibmodel}. We calculate average non-operational residence time in each shell using the mean cross-sectional area and mass of all intact objects from DISCOS. We assume fragments are uniform aluminium spheres with 10 cm diameter when calculating their residence time in each shell. We compute the rate at which objects transit from one shell to the next lower shell as the inverse residence time. We similarly obtain post-mission disposal (PMD) rates by operator type from DISCOS. 
Data on launch vehicle prices, sectoral revenues, and government budgets are used in the first and second-stage models. Launch vehicle price estimates are available for free from releases of the Federal Aviation Administration's (FAA's) Annual Compendium of Commercial Space Transportation \cite{faa}, while sectoral revenues and government budgets are available from The Space Report \cite{spacereport}. All data are recorded at an annual frequency or were converted to annual frequency.

Table \ref{tab:orbitdata} shows annual summary statistics of the physical object and launch traffic data used to estimate the first-stage model and validate the debris environment model. The object categories and classifications are described in section \ref{pibmodel}. Table \ref{tab:econdata} describes the economic data used to estimate both stages of the econometric model. The data for the first-stage model is structured as a set of ``choice occasions'', i.e. occasions where an operator chose to launch to one shell out of a collection of possible shells while observing the state of each shell. Each satellite successfully placed into orbit corresponds to one choice occasion.
The data contain 574 choice occasions for commercial operators (excluding amateur and large constellation operators), 1,171 choice occasions for civil government operators, and 463 choice occasions for defense operators over the 2007-2020 period.

The operator-type-specific launch vehicle prices (summarized in Table \ref{tab:econdata}) were calculated as the mean estimated price for launch vehicles used by an operator type in a particular sector in a given year. Thus, the price for a particular operator type in a particular year reflects both the types of missions that operator type conducts as well as the availability and capabilities of existing launch vehicles. The launch vehicle price data were incomplete, with 355 launch events having an estimated price while 2,898 did not. Missing launch vehicle prices were imputed using a random forest to model launch vehicle price as a function of launch vehicle, launch site, operator classification, orbital parameters, year, and month \cite{breiman2001forest, pantanowitz2009missing}. Summary statistics of observed and imputed launch vehicle prices are shown in Table \ref{tab:imputationstats}. 

The launch traffic, orbital state, and launch prices data all contain observations in the year 2020. However, the sectoral revenues and government budget data stop at 2019. The 2020 values of these data are imputed by applying a 15\% growth rate to all categories in 2019. This 15\% growth rate is consistent with projections in \cite{jonasetal} that the space economy relating to satellites would reach a size of 1.1 trillion USD by 2040. Other projections can be used to assess the sensitivity of the debris environment to sectoral revenues or national government spending, e.g. higher or lower insurance premiums or changes to government space budgets.

\begin{table}[!htbp] \centering 
  \caption{Physical object and launch traffic summary statistics. All values are counts.} 
\begin{tabular}{@{\extracolsep{5pt}}lcccc} 
\\[-1.8ex]\hline 
\hline \\[-1.8ex] 
& \multicolumn{1}{c}{Mean} & \multicolumn{1}{c}{St. Dev.} & \multicolumn{1}{c}{Min} & \multicolumn{1}{c}{Max} \\ 
\hline \\[-1.8ex] 
Civil & 215 & 72 & 121 & 338 \\ 
Active & & & & \\
Payload & & & & \\
 & & & & \\
Commercial & 313 & 199 & 165 & 703 \\ 
Active & & & & \\
Payload & & & & \\
 & & & & \\
Defense  & 129 & 50 & 81 & 227 \\ 
Active & & & & \\
Payload & & & & \\
 & & & & \\
Other & 82 & 79 & 14 & 250 \\ 
Active & & & & \\
Payload & & & & \\
 & & & & \\
Total & 740 & 395 & 398 & 1,518 \\ 
Active & & & & \\
Satellites & & & & \\
 & & & & \\
RB & 843 & 43 & 775 & 915 \\ 
& & & & \\
IP & 1,566 & 151 & 1,370 & 1,821 \\ 
& & & & \\
MRO & 666 & 76 & 570 & 814 \\ 
& & & & \\
COF & 8,271 & 1,050 & 5,856 & 9,122 \\ 
& & & & \\
Total & 11,348 & 1,261 & 8,571 & 12,581 \\ 
debris & & & & \\
& & & & \\
Commercial & 137 & 73 & 39 & 301 \\ 
launches & & & & \\
& & & & \\
Civil & 332 & 133 & 85 & 534 \\ 
government & & & & \\
launches  & & & & \\
 & & & & \\
Defense & 161 & 119 & 69 & 490 \\ 
launches & & & & \\
  & & & & \\
\hline \\[-1.8ex] 
\end{tabular} 
  \label{tab:orbitdata} 
\end{table} 


\begin{table}[!htbp] \centering 
  \caption{Economic data summary statistics. Launch vehicle prices are in million USD, while sectoral revenues and government budgets are in billion USD.} 
\begin{tabular}{@{\extracolsep{5pt}}lcccc} 
\\[-1.8ex]\hline 
\hline \\[-1.8ex] 
Statistic & \multicolumn{1}{c}{Mean} & \multicolumn{1}{c}{St. Dev.} & \multicolumn{1}{c}{Min} & \multicolumn{1}{c}{Max} \\ 
\hline \\[-1.8ex] 
Launch vehicle & 16.015 & 3.811 & 9.440 & 25.459 \\ 
prices & & & & \\
(Civil   & & & & \\
Government) & & & & \\
& & & & \\
Launch vehicle & 31.754 & 18.782 & 13.833 & 82.000 \\ 
prices & & & & \\
(Commercial)  & & & & \\
& & & & \\
Launch vehicle & 17.079 & 3.059 & 11.900 & 22.333 \\ 
prices & & & & \\
(Defense)  & & & & \\
& & & & \\
Insurance & 0.753 & 0.175 & 0.460 & 0.982 \\ 
premiums & & & & \\
& & & & \\
Commercial & 2.012 & 0.482 & 1.210 & 2.613 \\
satellite & & & & \\
launch & & & & \\ 
& & & & \\
Commercial & 4.761 & 1.028 & 3.410 & 6.817 \\ 
satellite & & & & \\
manufacturing & & & & \\
& & & & \\
Direct-to-Home & 86.197 & 13.522 & 55.400 & 97.800 \\ 
TV & & & & \\
& & & & \\
Satellite & 20.874 & 3.190 & 15.100 & 24.970 \\ 
communications & & & & \\
& & & & \\
Satellite & 4.354 & 1.884 & 2.100 & 8.024 \\ 
radio & & & & \\
& & & & \\
Earth & 2.483 & 0.825 & 1.268 & 3.687 \\ 
observation & & & & \\
& & & & \\
Infrastructure & 123.289 & 53.251 & 61.713 & 210.923 \\ 
& & & & \\
US government & 49.658 & 8.136 & 39.301 & 64.674 \\ 
& & & & \\
Non-US & 31.978 & 6.868 & 17.962 & 40.934 \\ 
governments & & & & \\
& & & & \\
\hline \\[-1.8ex] 
\end{tabular} 
  \label{tab:econdata} 
\end{table} 

We calibrate the physical model using data from the European Space Agency regarding launch traffic and satellite and debris stocks in the 600-650km shell.

\begin{table}[!htbp] \centering 
  \caption{Summary statistics of observed and imputed launch prices. All values shown are in million USD.} 
\begin{tabular}{@{\extracolsep{5pt}}lccccc} 
\\[-1.8ex]\hline 
\hline \\[-1.8ex] 
& \multicolumn{1}{c}{N} & \multicolumn{1}{c}{Mean} & \multicolumn{1}{c}{St. Dev.} & \multicolumn{1}{c}{Min} & \multicolumn{1}{c}{Max} \\ 
\hline \\[-1.8ex] 
Observed & 355 & 17.13 & 7.83 & 1 & 33 \\ 
 & & & & \\
Imputed & 2,898 & 20.28 & 6.35 & 5 & 33 \\ 
 & & & & \\
Combined  & 3,253 & 19.94 & 6.60 & 1 & 33 \\ 
  & & & & \\
\hline \\[-1.8ex] 
\end{tabular} 
  \label{tab:imputationstats} 
\end{table} 

\subsection{Debris environment model}
\label{pibmodel}
We use a Particle-in-a-Box (PIB) model of the debris environment similar to those used in \cite{lewis2009fade, somma2017statistical}. A key innovation in this PIB model is that active satellites are distinguished by the type of operator who owns them, facilitating linkage with the choice model. The PIB model segments the low-Earth orbit (LEO) region from 100-1300 km altitude into 50 km spherical shells, and distinguishes tracked objects within the shells by their physical and economic characteristics. Debris objects are segmented into four classes: rocket bodies (RB), mission-related objects (MRO), intact inactive payloads (IP), and miscellaneous debris and fragments (COF). Satellites are segmented into 5 classes for the debris environment model: commercial, civil government, defense, amateur, and large constellations. The amateur operators category includes non-governmental operators who do not have commercial motives. Separating large constellation satellites enables using pre-announced constellation launch plans in projections, while separating amateur operators' satellites ensures that the commercial category appropriately represents operators with commercial motives while estimating the econometric model.

The governing equations of the debris environment model are shown in Equations \ref{sateqn} and \ref{debeqnIP}-\ref{debeqnMRO}. The model has a total of 24 shells (with $j$ being the shell immediately above $j-1$), 9 object species (4 debris types and 5 operator types), and operates at an annual timestep. $S_{ij}$ refers to the current stock of active satellites from operator type $i$ in shell $j$ and $D_{kj}$ refers to the current stock of debris objects of type $k$ currently in shell $j$. $\delta_j$ refers to the annual rate of orbital decay in shell $j$ to shell $j-1$, $\mu_{i}$ refers to the rate at which active satellites reach the end of their productive lives and transition to being intact inactive payloads, and $L_{ij}$ or $L_{kj}$ refer to the rate of ``unavoidable'' collisions involving objects of type $i$ or $k$ (i.e. the number of objects removed due to collisions).\footnote{While only active satellites can actively avoid collisions, PIB models often feature adjustment coefficients which are calibrated to match higher-fidelity debris environment models, e.g. \cite{sommathesis}. The implementation of active collision avoidance and these calibration parameters in this model is algebraically identical. Consequently both $L_{ij}$ and $L_{kj}$ are referred to as ``unavoidable'' collisions to reflect that both contain coefficients which adjust them from a kinetic gas model of collision rates, though the coefficients carry different interpretations.} $q_{ij}$ is the number of satellites launched by operator type $i$ to shell $j$, $\gamma_{kj}$ is the number of fragments created in shell $j$ when an object of type $k$ is destroyed, $\rho_{ij}$ is the number of rocket bodies created by launching a single satellite of type $i$ to shell $j$, and $\mu^s_{ij}$ and $\mu^l_{ij}$ are the numbers of mission-related objects created annually by a single satellite and a single launch of type $i$ in shell $j$. All active satellites are assumed to be perfectly coordinated and avoid all collisions. 
The rates at which all species of debris objects collide with each other are adjusted to 0.01\% of the rates predicted by the PIB model. 
Future work calibrating this PIB model to long-term simulations from higher-fidelity debris environment models will produce more realistic collision and debris accumulation patterns.

PMD is modeled by having a fraction of satellites reaching their end-of-life each year move to an orbit where they will decay in 25 years to comply with the 25-year guideline \cite{IADC2020} (not shown in Equations \ref{debeqnIP}-\ref{debeqnMRO}). For intact inactive payloads in our model, the 500-550 km shell is the highest one with a natural decay time under 25 years. Only satellites above this altitude are moved when they reach their end-of-life---satellites which are already below this level are naturally compliant with the 25-year guideline.

\begin{eqnarray}
 \dot{S}_{ij}&=& q_{ij} - (\mu_i - L_{ij})S_{ij} \label{sateqn} \\
 \dot{D}_{IP,j} &=& \sum_i \mu_{i} S_{ij} - L_{IP,j} - \label{debeqnIP} \\ 
 &~& \delta_{IP,j} D_{IP,j} + \delta_{IP,j+1} D_{IP,j+1} \nonumber \\
 \dot{D}_{COF,j} &=& \sum_k \gamma_{kj} L_{kj} + \sum_i \gamma_{ij} L_{ij} - \label{debeqnCOF} \\ 
 &~& \delta_{COF,j} D_{COF,j} + \delta_{COF,j+1} D_{COF,j+1} \nonumber \\
 \dot{D}_{RB,j}&=& \sum_i \rho_{ij} q_{ij} - L_{RB,j} - \label{debeqnRB}\\
 &~& \delta_{RB,j} D_{RB,j} + \delta_{RB,j+1} D_{RB,j+1} \nonumber \\
 \dot{D}_{MRO,j}&=& \sum_i (\mu^s_{ij} S_{ij} + \mu^l_{ij} q_{ij}) - L_{MRO,j} - \label{debeqnMRO} \\
 &~& \delta_{MRO,j} D_{MRO,j} + \delta_{MRO,j+1} D_{MRO,j+1} \nonumber
\end{eqnarray}

For simplicity we set $\rho_{ij} = \mu^s_{ij} = \mu^l_{ij} = 0 ~ \forall i, j$, so that no new rocket bodies or mission-related objects are created during simulations. This implies that all rocket bodies are perfectly disposed of after insertion and that no mission-related objects are released. The actual current disposal rate and release rate for these two classes of objects can be found in \cite{envRep2020}.  Collision rates are calculated following \cite{sommathesis} (Equations 2.21-2.23) and fragmentation rates are calculated following the NASA standard breakup model \cite{krisko2011}.

\subsection{Econometric model of satellite launching}

Multi-stage budgeting approaches to consumer behavior have a long history in economics \cite{deaton1980aids, jorgensen1988budgeting, hausman1995utility, nevo2011models}. The idea is to consider the consumer's decision problem (``how to allocate resources across a variety of goods'') as a sequence of separate-but-related decision problems. Such separation facilitates modeling consumer demands flexibly while retaining desirable demand aggregation properties and keeping the dimensionality of the system to be solved manageable \cite{nevo2011models}. One popular approach is to use two stages: the choice of how much to purchase in total (e.g. total expenditure, total number of trips) and the choice of how to allocate total purchases across the available products (e.g. how much of each good to buy, how frequently to visit each site). Such models have been applied in a wide variety of settings, from consumer demands for a broad basket of goods \cite{deaton1980aids}, to demands for specific goods such as energy \cite{jorgensen1988budgeting}, and even to unpriced goods such as recreational trips to beaches \cite{hausman1995utility}. In some cases, e.g. \cite{hausman1995utility}, the model can be used to estimate damages due to disasters and to calculate the appropriate level and scope of regulations to deter future environmental damages.

Our two-stage model of orbital demand (i.e. satellite launching) involves a representative operator of each modeled type (commercial, civil government, and defense) first choosing the total number of satellites to launch and then choosing how to allocate those satellites over orbital shells. We first estimate the second-stage model to generate choice probabilities over shells, and then calculate the choice-probability-weighted mean utility level across orbital shells. This weighted mean utility is then used as a predictor of the representative operator's choice of how many satellites to launch in the first-stage model, similar to the price index approach in \cite{hausman1995utility}. It is shown in \cite{hausman1995utility} that estimating the second stage first and using its output to inform the first stage ensures the choices at each stage are consistent with each other and with utility-maximizing behavior.

\subsubsection{The second stage: choice over orbits}
\label{econometric_stage_2}
Each representative operator allocates $N_{it}$ satellites across $J$ orbits in period $t$. We analyze this decision using a multinomial logit model of random utility maximization, e.g. \cite{hausman1995utility, train1998recreation}. These ``random utility models'' (RUMs) posit that a decision maker faced with a set of discrete alternatives (``products'') and the opportunity to select one will select the product which maximizes a scalar-valued function (their ``utility''). The decision maker's utility function can depend on their own attributes and on attributes of the products, and the researcher may only observe some of the relevant attributes. The utility function for an operator of type $i$ considering launching to shell $j$ at time $t$ is shown in Equation \ref{utility}.

\begin{equation}
    U_{ijt} = \alpha_{ij} + X_{jt} \beta_i + ac_{ijt} \gamma_i + \nu_{ijt}. \label{utility}
\end{equation}

In Equation \ref{utility}, $\alpha_{ij}, \beta_i, \gamma_i$ are parameters to estimate. $X_{jt}$ is a vector of characteristics of orbit $j$ at time $t$, including the distributions of satellites and debris across operator and object types and the pre-avoidance collision rate predicted by the PIB model in shell $j$, and $\beta_i$ are the associated marginal utilities of the characteristics. $ac_{ijt}$ is the estimated cost of accessing shell $j$ for operator type $i$ in time $t$, calculated from the imputed launch vehicle price estimates in each year and the minimum delta-v required to reach the orbit. $\gamma_i$ is an estimate of the marginal utility of money, and can be used for valuation exercises. The access cost is measured in units of millions of \$-GJ (dollar-gigajoules).\footnote{The access cost for an operator of type $i$ launching to shell $j$ in year $t$ is the product of the launch vehicle price faced by $i$ at $t$ and the minimum delta-v required to reach shell $j$. This is similar to approaches used to construct travel costs in recreational demand models, e.g. \cite{timmins2007revealed}. Access costs can be converted from \$-GJ to dollar units using launch vehicle lift capacities. While this unit conversion will facilitate valuation and tax policy analysis, it is not necessary for the counterfactual scenarios we consider here.} $\alpha_{ij}$ is an orbit-operator fixed effect or ``alternative-specific constant'' which captures idiosyncratic operator preferences (including unmodeled factors) over orbits.\footnote{Alternative-specific constants (ASCs) can be interpreted as the mean utility of product $j$ when all of the other observable components are set to zero. In general, with $J$ alternatives only $J-1$ ASCs can enter the model, reflecting that only differences in utility matter for observed choice outcomes (i.e. the scale of utility is irrelevant). ASCs are frequently decomposed into functions of observed product characteristics and a separate unobserved error term for welfare analysis, e.g. assessment of damages as in \cite{timmins2007revealed}. We forgo this step here as we focus on short-run counterfactual simulation rather than welfare analysis.} $\nu_{ijt}$ is an error term representing factors influencing the operator's choice but unobserved to the researcher.

The debris environment model distinguishes between amateur operator satellites and large constellation satellites to better account for their different physical characteristics. However, the amateur and large constellation categories are collapsed into a  single category, ``other'', when estimating the second-stage behavioral model. Since the decisions to launch these satellites are not being explicitly modeled, this grouping streamlines the simulation and reduces one parameter to estimate in this stage. The coefficient associated with the total number of satellites belonging to ``other'' operators should therefore be interpreted as a representative operator's marginal utility from increases in the population of amateur or large constellation operators.

Our assumption that the decision maker maximizes their utility, combined with the unobserved random error term $\nu_{ijt}$, implies that the probability a decision maker of type $i$ selects orbit $j$ in period $t$ (the ``choice probability'') is as shown in Equation \ref{choiceprob}.

\begin{equation}
    P_{ijt} \equiv Pr(U_{ijt} > U_{ikt} ~ \forall k) \label{choiceprob}
\end{equation}

Assuming that all $\nu_{ijt}$ are independently and identically distributed according to a Type 1 Extreme Value distribution produces a multinomial logit model of $P_{ijt}$, shown in Equation \ref{CP_logit}.\footnote{Other distributional assumptions yield different models of $P_{ijt}$, e.g. assuming the unobserved components of utility are normally distributed yields a probit model as in \cite{bolduc1996doctor}.} 

\begin{equation}
    P_{ijt} = \frac{\exp(\alpha_{ij} + X_{jt} \beta_i + ac_{ijt} \gamma_i)}{\sum_{k=1}^J \exp(\alpha_{ik} + X_{kt} \beta_i + ac_{ikt} \gamma_i)}. \label{CP_logit}
\end{equation}

The total number of shells is represented in Equation \ref{CP_logit} by $J$ for generality. The multinomial logit specification is widely used due to its computational simplicity (e.g. \cite{hausman1995utility, train1998recreation, timmins2007revealed}), although it imposes some restrictions on the substitution patterns the model can produce \cite{trainbook}. We discuss limitations of the multinomial logit model in Section \ref{limitsandfuture}, along with some potential solutions. We estimate the model parameters $\alpha_{ij}, \beta_i, \gamma_i$ by maximizing the log likelihood of the observed choice shares over orbits using data from 2007-2020. We estimate Equation \ref{CP_logit} separately for commercial, civil government, and defense operators to obtain operator-specific coefficients. We model ``other'' operators using repeated launch patterns, as these operators are a diverse-enough user class that a single multinomial logit model will likely approximate their behaviors poorly. 

We show estimates of $\beta_i$ and $\gamma_i$ for commercial, civil government, and defense operators in Table \ref{tab:rum_estimates}. The parameter estimate magnitudes reveal the expected change in the log odds of an operator type launching to a particular shell for a one-unit increase in the variable, e.g. a 1 million \$-GJ increase in the cost of accessing a particular shell is associated with a 0.17 decrease in the log odds that a commercial operator will launch to that shell. Exponentiating the estimates yields the equivalent odds ratio, i.e. an estimate of -0.017 implies an odds ratio of around 0.98, suggesting that a 1 million \$-GJ increase in accessing a particular shell is associated with a 2\% reduction in the chance a commercial launches there and corresponding increases in the chance they launch elsewhere.  Standard errors are shown in parentheses, though we do not attempt any statistical inference on the estimates. More research is required to determine the most appropriate structures of the behavioral model and covariance matrix of parameter estimates.

\begin{table}[!htbp] \centering 
  \caption{Second-stage parameter estimates} \vspace{1em}
    \renewcommand{\arraystretch}{0.9}
\begin{tabular}{@{\extracolsep{1pt}}lccc} 
\\[-1.8ex]\hline 
\hline \\[-1.8ex] 
 & \multicolumn{3}{c}{\textit{Choice of orbit}} \\ 
\cline{2-4} 
\\[-1.8ex] & Commercial & Civil & Defense\\ 
\\[-1.8ex] &  & Government & \\ 
\hline \\[-1.8ex] 
 Civil& $-$0.006 & 0.017 & 0.065  \\ 
  Active Payload & (0.010) & (0.007) & (0.013) \\ 
  & & & \\ 
 Commercial & 0.015  & 0.016  & 0.014  \\ 
  Active Payload & (0.002) & (0.002) & (0.003) \\ 
  & & & \\ 
 Defense & $-$0.021 & 0.003 & 0.011 \\ 
  Active Payload& (0.012) & (0.008) & (0.013) \\ 
  & & & \\ 
 Other & $-$0.001 & 0.011  & 0.053  \\ 
 Active Payload & (0.004) & (0.003) & (0.007) \\ 
  & & & \\ 
 Access cost & $-$0.017  & $-$0.019 & $-$0.022 \\ 
  & (0.005) & (0.010) & (0.020) \\ 
  & & & \\ 
 Total PIB & $-$0.003 & $-$0.013  & $-$0.052  \\ 
 collision rate & (0.004) & (0.003) & (0.007) \\ 
  & & & \\ 
\hline \\[-1.8ex] 
Observations & 574 & 1,171 & 463 \\ 
Log Likelihood & $-$1,210 & $-$2,833 & $-$963 \\ 
\hline 
\hline \\[-1.8ex] 
\end{tabular} 
  \label{tab:rum_estimates} 
\end{table} 

The parameter estimates indicate that commercial, civil government, and defense operators all reduce their launch activity to orbits as access costs and collision likelihood increase. This implies that all operator types will respond to reductions in the cost of accessing a particular region by launching more satellites there, and to increases in collision likelihood in a region by launching fewer satellites there. The variable ``Total PIB collision rate'' measures the total number of collisions predicted by the debris environment model before applying any collision avoidance or adjustment coefficients. The variable is constructed to include all possible collisions, as avoidance will likely be costly and the total number of predicted collisions will proxy for the cost of avoiding collisions. Another way to understand the relationship between the estimates for access costs and collision rates is to consider a change to both cost and collision rate which would leave utility and choice probabilities unchanged. The parameter estimates suggest that operators would accept additional collision risk only if it came with cost savings, or vice versa. These estimates reflect the fundamental relationship derived in \cite{adilov2018kessler}, \cite{raoetal2020} and \cite{rouillon2020economic} between payoffs and risks when deciding to launch a satellite, and suggests that the second-stage model is consistent with economic theories of orbit use by rational utility maximizers.

Interpreting the parameters for active payload variables is more challenging. To the extent that the total collision rate predicted by the debris environment model proxies for the total cost of avoiding collisions, the parameters on active payloads do not reflect behavioral responses to potential collision risk posed by operators' satellites. Idiosyncratic operator-type-specific preferences for specific orbital shells are captured in the alternative-specific constants $\alpha_{ij}$. However, unmodeled factors may be correlated with both the decision to launch to a shell and the number of satellites of a given type already there. Such factors might include the revenues or payoffs received by each operator type from satellites in a particular shell, availability of radio spectrum, and plans to build new constellations. These types of factors are currently only captured by the current presence of satellites from different operator types. Consequently, positive parameter estimates for satellites owned by different operator types may reflect common underlying incentives to use or avoid an orbit based on unmodeled factors or may represent direct behavioral responses to the presence of other operators. This reflects a broader challenge with estimating behavioral parameters in the presence of congestion effects or strategic motives. As discussed in \cite{timmins2007revealed}, such cases may require more sophisticated estimation approaches, e.g. the use of instrumental variables, to appropriately capture responses to congestion or other strategic responses. Measuring and distinguishing between such effects is an important area for future work.

\subsubsection{The first stage: number of satellites launched}

In the first stage, each representative operator chooses a number of satellites $N_{it}$ to launch each period, recognizing that they will be allocated across orbits according to the second-stage model. Operators' decisions in this stage are based on sectoral revenues across the space industry, government space budgets, the average collision rate expected to occur in the absence of any avoidance maneuvers or adjustments, and a price index for orbit use reflecting access costs, orbit quality, and other factors from the first stage. We use the choice-probability-weighted utility from the first stage as our price index in the second stage. Since the number of satellites launched is a non-negative integer, a count regression model is appropriate \cite{hausman1984models, cameron1986models}. This approach specifies the conditional expectation of the number of satellites launched by a representative operator of type $i$ in period $t$ as shown in Equation \ref{countmodel}.

\begin{equation}
    E[N_{it} | Z_{it}] = \exp(Z_{it} \omega_i). \label{countmodel}
\end{equation}

In Equation \ref{countmodel}, $Z_{it}$ contains variables which affect the demand for launching satellites (including an intercept term and the price index derived from the first stage). Calculating a quasi-maximum likelihood estimate (QMLE) of $\omega_i$ implies that $N_{it}$ is a Poisson random variable with conditional mean as shown in Equation \ref{countmodel}. One advantage of this approach is that as long as the conditional mean is correctly specified, the QMLE of $\omega_i$ is consistent. We estimate Equation \ref{countmodel} separately for commercial, civil government, and defense operators to obtain operator-specific coefficients.

We estimate $\omega_i$ using data on sectoral revenues and government budgets, summarized in Table \ref{tab:econdata}, as well as the unadjusted collision rate calculated from the debris environment model and the weighted mean utility calculated from the first stage. Using these data poses a challenge: the data from 2007-2020 gives $n=14$ observations, while there are $p=13$ variables in $Z_{it}$. Though the regression can be estimated when $n$ is close to $p$, the variance of the estimates will be high and the regression will likely have poor out-of-sample performance. \footnote{One solution is to reduce the number of variables by either discarding some or by combining the variables in a theoretically-appropriate fashion. Discarding or combining variables without theory risks rendering the estimates inconsistent, while the appropriate theoretical restrictions are not obvious. The economic theory of satellite launching decisions is still in its infancy, with relatively few studies analyzing the determinants of commercial satellite launching decisions (e.g. \cite{adilov2015economic, rouillon2020economic, raoetal2020}) or of commercial orbital debris removal (e.g. \cite{muller2017valuation, brettle2019future, brettle2020commercial}). The majority of these studies are theoretical and do not conduct statistical tests to determine the appropriateness of different model specifications. Though \cite{raoetal2020} derives appropriate variable combinations for commercial operators from an underlying theoretical economic model, we were unable to find similar analyses of civil government or defense launch decisions.} 

To avoid ad hoc variable selection, we apply a ridge penalty to estimate Equation \ref{countmodel} while controlling the variance of the estimates and including all variables. Ridge regression is a well-known technique for estimating regression models where $p \approx n$ or even $p > n$ \cite{hoerl1985ridge, dobriban2018ridge}. Intuitively, ridge regression uses the bias-variance tradeoff to obtain more precise estimates and better out-of-sample predictive performance at the expense of allowing biased (though still consistent) estimates. Ridge regression achieves this performance by introducing a tunable quadratic penalty term in the estimation objective function, ensuring that parameter values are ``shrunk'' (i.e. biased) toward zero without also performing variable selection.\footnote{While using biased regressions limits our ability to conduct statistical inference on estimated parameters, our limited sample size already makes such inference impossible.} We estimate the tunable parameter of the penalty term using $k$-fold cross-validation. Simulation analysis of Poisson ridge regression suggests it performs well in controlling parameter instability \cite{mansson2011poisson}. We discuss other ways to address this sample size issue in Section \ref{limitsandfuture}. We show estimates of $\omega_i$ for commercial, civil government, and defense operators in Table \ref{tab:count_estimates}. 

\begin{table}[!htbp] \centering 
  \caption{First-stage parameter estimates} 
\begin{tabular}{@{\extracolsep{1pt}} ccccc} 
\\[-1.8ex]\hline 
\hline \\[-1.8ex] 
 & \multicolumn{3}{c}{\textit{Number of satellites launched}} \\ 
\cline{2-4} 
 & Commercial & Civil & Defense \\ 
 & & Government & \\ 
\hline \\[-1.8ex] 
Intercept & 8.677 & 5.701 & 3.716 \\ 
& & & \\ 
Insurance & $-$3.917 & 0.100 & $-$0.283 \\ 
premiums &  &  & \\ 
& & & \\ 
Commercial & $-$0.231 & $-$0.262 & 0.347 \\ 
satellite &  &  & \\ 
launch &  &  & \\ 
& & & \\ 
Commercial & $0.335$ & 0.009 & 0.238\\ 
satellite  & & &\\ 
manufacturing  \\ 
& & & \\ 
Direct-to-home& $-$0.003 & $-$0.004 & 0.011 \\ 
TV & & & \\ 
& & & \\ 
Satellite & $-$0.097 & $-$0.004 & 0.001 \\ 
communications & & & \\ 
& & & \\ 
Satellite & $-$0.253 & $-$0.075 & $-$0.036 \\ 
radio & & & \\ 
& & & \\ 
Earth & $-$0.184 & $-$0.044 & $-$0.175 \\ 
observation & & & \\ 
& & & \\ 
Infrastructure & $0.002$ & $0.001$ & $0.0002$ \\ 
& & & \\ 
US government & $0.002$ & $-$0.007 & $-$0.035 \\ 
& & & \\ 
Non-US & 0.047 & $0.057$ & 0.034 \\ 
governments & & & \\ 
& & & \\ 
First-stage & 1.002 & $-$4.425 & $-$2.321 \\ 
mean utility & & & \\ 
& & & \\ 
Mean total PIB & $-$0.001 & $-$0.007 & $-$0.004 \\ 
collision rate & & & \\ 
& & & \\
\hline \\[-1.8ex] 
\end{tabular} 
  \label{tab:count_estimates} 
\end{table} 

Interpreting the parameter estimates in Table \ref{tab:count_estimates} is more challenging than interpreting those in Table \ref{tab:rum_estimates} due to the bias induced by the ridge regression and the inherent uncertainty from the small sample size. To the extent that we can make any inferences about the determinants of launch behavior, the signs of the estimates may give some useful information. In particular, the sign on the parameter for ``Mean total PIB collision rate'' gives some indication as to how the total number of satellites launched will respond to a measure of the average collision likelihood (based on a projection from the PIB model). The negative sign on this estimate across all operator types indicates that operators will tend to launch fewer satellites as the orbital environment becomes riskier. The utility term can be interpreted as a price index for orbit use, similar to the one used in \cite{hausman1995utility}, since the expected utility from a logit regression has the same form as consumer surplus.

Incorporating a measure of the expected collision rate into the first stage, along with the expected total utility across all shells, may be consistent with ``imperfect'' open access behavior. This behavior is modeled in \cite{raoetal2020} using the total launch rate and an ``exact'' open access condition, albeit in a model where all of the 100-2000 km altitude region is treated as a single shell and there is no need for a second stage with choice over shells. The ``exact'' open access condition implies there are no errors in estimating the net payoffs from launching to a particular shell in a particular year. 
Using the mean collision rate and mean utility over all shells rather than the full distribution of collision rates and utilities allows open access to hold only in expectation (i.e. imperfectly) across shells.
This allows satellites belonging to specific operator types in specific shells to produce positive or negative net payoffs, with individuals within a user class arbitraging away type-specific net payoffs by sorting themselves across shells. There may be other interpretations consistent with this specification; identifying the most appropriate first-stage model specifications and interpretations is an important area for future work.

\section{Model validation}

This section presents preliminary validation of the integrated model, comparing outputs from the integrated model to observed data. There remain a number of margins along which the model must be improved. These include a detailed calibration of the PIB model along the lines of the calibration conducted in \cite{sommathesis}, more flexibility in the second-stage model, and more temporal disaggregation of the first-stage model.

\begin{figure}[h!]
\centering
\includegraphics[width=\linewidth]{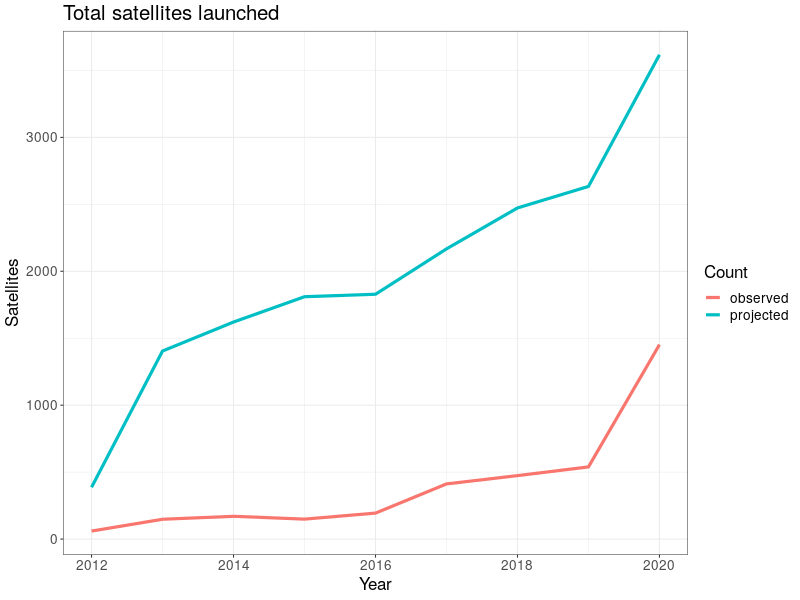} 
\caption{Projected aggregate satellite stock from integrated model compared to observed data.} \label{fig:model_validation1}
\end{figure}

\begin{figure}[h!]
\centering
\includegraphics[width=\linewidth]{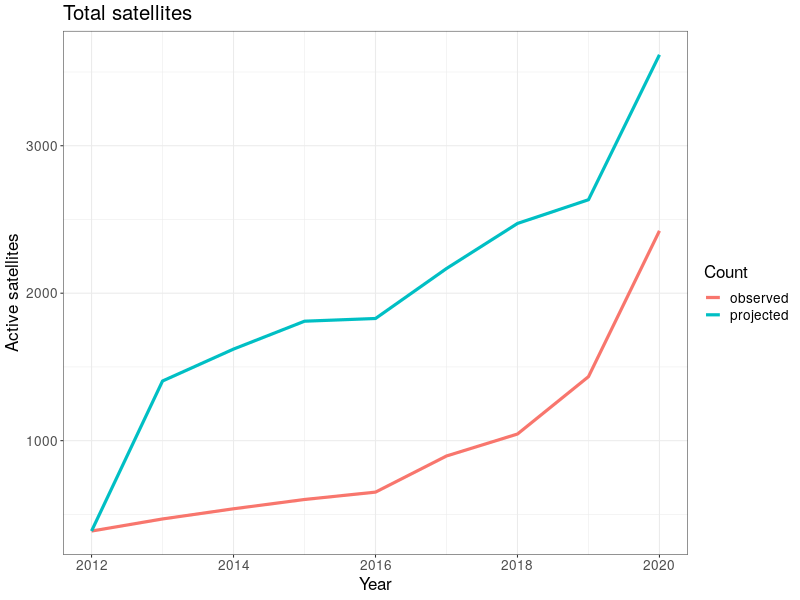} 
\caption{Projected aggregate satellite stock from integrated model compared to observed data.} \label{fig:model_validation2}
\end{figure}

\begin{figure}[h!]
\centering
\includegraphics[width=\linewidth]{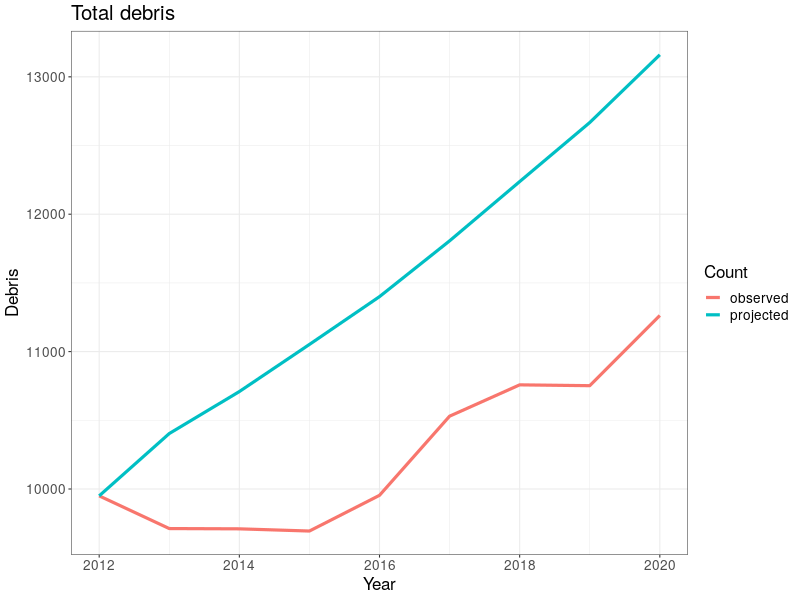} 
\caption{Projected aggregate debris stock from integrated model compared to observed data.} \label{fig:model_validation3}
\end{figure}

Figures \ref{fig:model_validation1}, \ref{fig:model_validation2}, and \ref{fig:model_validation3} show the aggregate launch flows, satellite stocks, and debris stocks projected from the integrated model. The projections begin at the 2012 orbital state and evolve till 2020. Although the launch flows are overprojected, the projected launch flows and satellite stock move in the correct direction, increasing with sectoral revenues and government budgets. Estimating the first-stage model with more observations (e.g. disaggregating to the monthly frequency) may reduce the extent to which it overprojects launch rates. The projected debris stock grows faster than the observed series, suggesting that calibrating the PIB model to long-run outcomes from a higher-fidelity debris environment model is a necessary step to maturing this framework.
 

\begin{figure*}
\centering
\includegraphics[width=\linewidth]{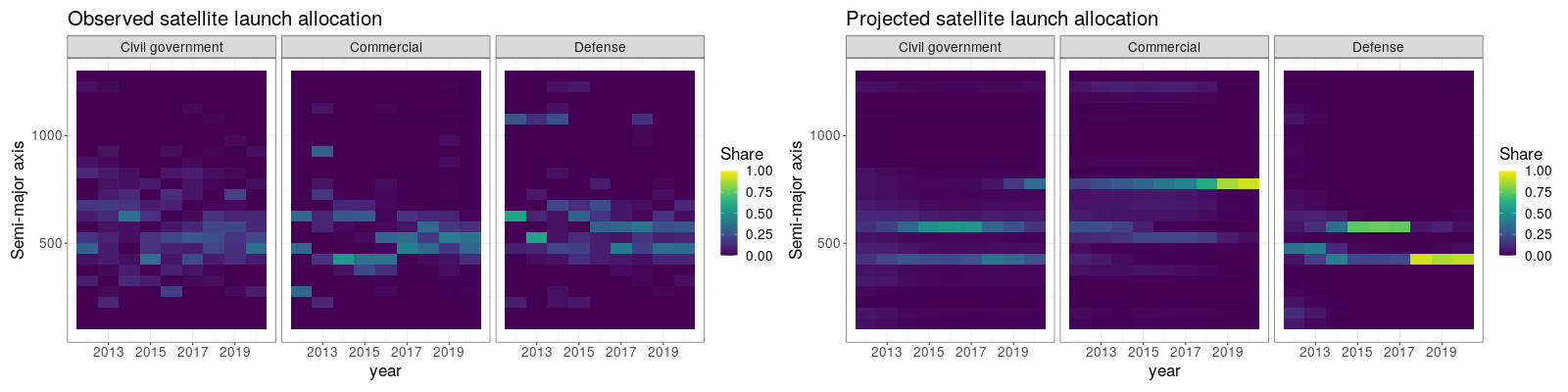} 
\includegraphics[width=\linewidth]{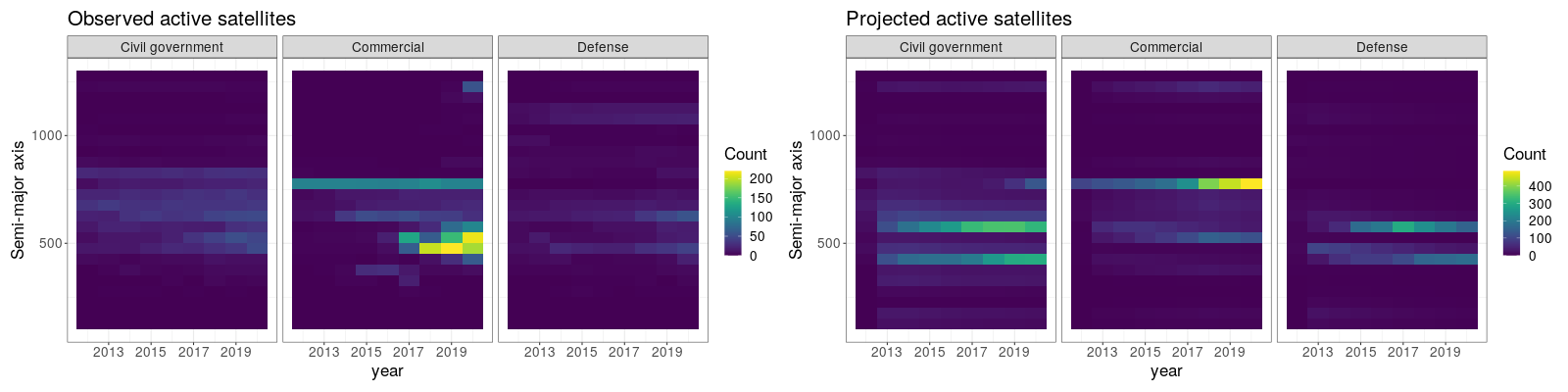} 
\includegraphics[width=\linewidth]{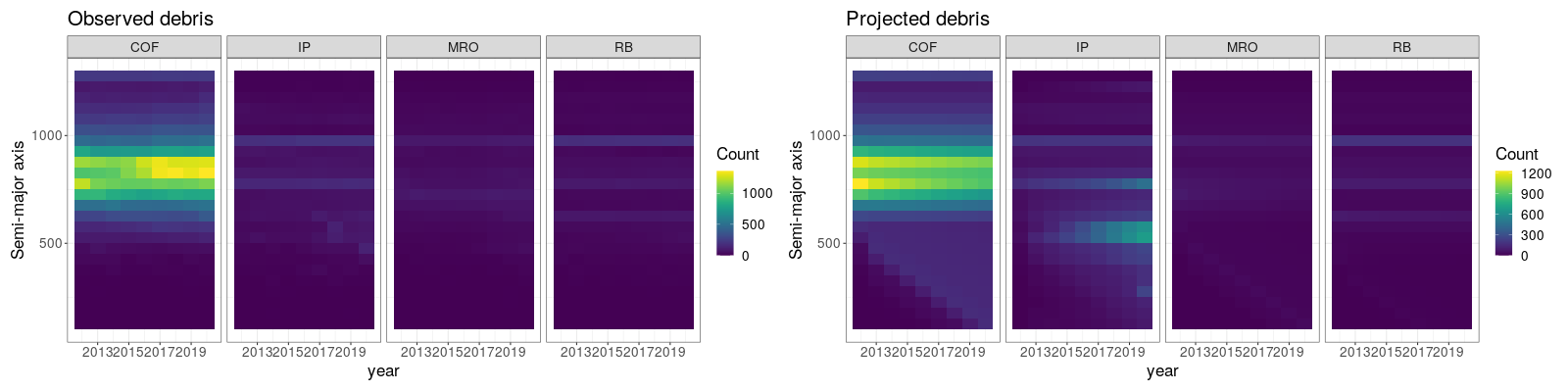} 
\caption{Observed and projected spatial distribution of satellites across modeled operator types and debris species.} \label{fig:model_validation4}
\end{figure*}

Figure \ref{fig:model_validation4} shows the projected and observed spatial distributions of launches and satellites by each modeled operator type and debris species. The satellite and debris projections capture some features of the observed data, notably the ``clustering'' of satellites in particular altitude shells and the qualitative debris decay patterns. However, there remains considerable scope for improvement. The projected launch allocations are more concentrated than observed patterns. While the model captures satellite clustering by different operator classes near 500 km, the projected magnitudes are too high (as shown in figures \ref{fig:model_validation1} and \ref{fig:model_validation2}). The commercial sector projection also does not capture the emergence of clustering between 475 and 525 km, instead projecting continued growth in the cluster at 775 km. The debris distribution reinforces the need for calibrating the PIB model to long-term trends. 

\section{Scenario analyses}

The scenarios considered here are illustrative of the types of analyses which can be conducted using this framework. Further work is required (e.g. calibrating the PIB model to long-term debris environment models) before using the framework for policy or mission-related decisions. The key message of this section is to highlight that integrated models such as this one reveal additional channels for the consequences of events or policies. 

The first set of scenarios show substitution responses across shells in response to unexpected fragmentations. The second scenario shows substitution responses to reduced access costs to lower altitudes and the corresponding changes in intact inactive payload accumulation. The third scenario shows substitution responses under increasing PMD compliance with the 25-year rule and the corresponding changes in intact inactive payload accumulation.

\subsection{Unexpected fragmentations}

\begin{figure*}
\centering
\includegraphics[width=\linewidth]{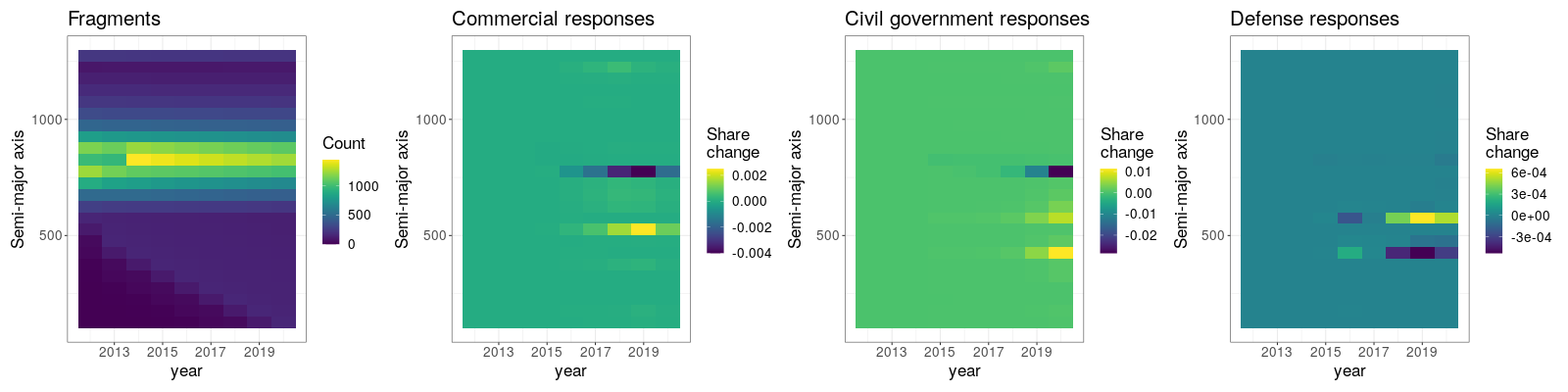} 
\includegraphics[width=\linewidth]{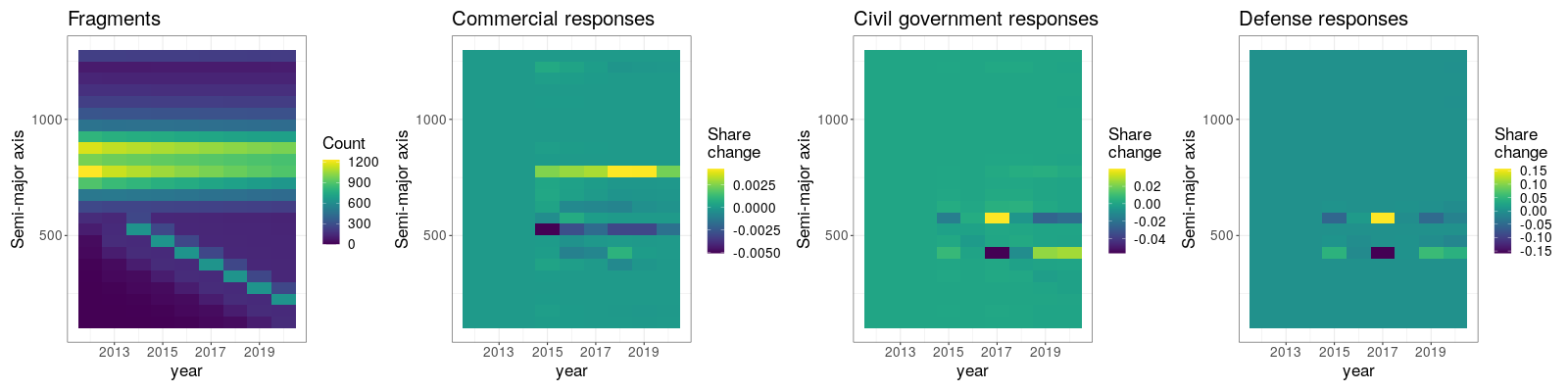}
\caption{Projected behavioral responses to a fragmentation event at different altitudes. The upper row shows an event occurring at 825-875 km altitude, while the lower row shows an event occurring at 525-575 km altitude.}  \label{fig:frag_events}
\end{figure*}

The fragmentation scenarios feature two separate events: one at a high altitude where debris will persist for years, and one at a low altitude where debris will quickly decay. Both the high and low altitude fragmentation events involve a total of 625 new fragments distributed over two shells, with 500 fragments in the lower shell and 125 fragments in the upper shell. The high-altitude fragmentation event occurs at 825 and 875 km altitudes and the low-altitude fragmentation event occurs at 525 and 575 km altitudes. The scenario begins at the orbital environment state in 2012 and runs till 2020, with the ``other'' operator type playing out its true historical launch behavior over the period. The fragments are initialized at the center of their orbital shells, and both occur at the end of simulation year 2014.  Figure \ref{fig:frag_events} shows the projected distribution of fragments and the second stage model responses for commercial, civil government, and military operators. Focusing on the second stage reveals the differing intensities of substitution responses across shells on a common scale.

Responses to both events reflect the aversion to collisions shown in Table \ref{tab:rum_estimates}. In the high-altitude fragmentation scenario operators tend to shift their launch allocation to lower shells, while in the low-altitude fragmentation scenario operators shift their launch allocations upwards. The defense sector response to the high-altitude fragmentation tends to shift to a higher shell, though the magnitude is small compared to responses from other sectors. High altitude debris persists longer than low-altitude debris, and induces longer-lasting substitution across shells. The response strengths across operator types tend to increase with the magnitude of the estimated ``Total PIB collision rate'' parameter (Table \ref{tab:rum_estimates} and the pre-event choice probabilities over shells (upper row of figure \ref{fig:model_validation3}). Intuitively, operators who are more averse to collisions will respond more to an increase in the fragment population. Those responses will be more pronounced when the collision rate increases in a shell the operator was otherwise likely to choose.

\subsection{Lower access costs}

In this scenario, altitudes below 500 km become comparatively cheaper to access, consistent with a rapid increase in small-rocket launch companies.  The scenario begins at the orbital environment state in 2018 and runs till 2030, with the ``other'' category repeating its 2020 launch pattern each year. Sectoral revenues and government budgets are projected forward at a 15\% annual growth rate to match the projections described in section \ref{data}. The costs of all launch vehicles used by all operator types decrease at an annual rate of 1\% after 2020 to reflect incremental launch sector productivity gains. In 2024, altitudes below 500 km experience an additional persistent 50\% access cost reduction.

\begin{figure}
\centering
\includegraphics[width=\linewidth]{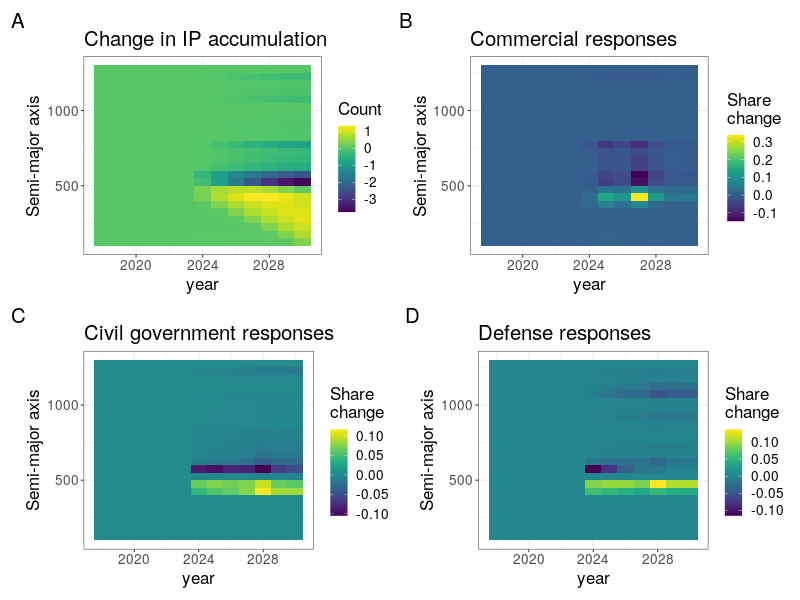}
\caption{Projected behavioral responses to a sharp decrease in the cost of accessing altitudes below 500 km.}  \label{fig:supply_shock}
\end{figure}

Responses to the cost reduction reflect the preferences for lower access costs shown in Table \ref{tab:rum_estimates}. All operator types take advantage of the reduction by shifting launches planned for the 500-550 km shell to the 400-500 km shells, and commercial operators are drawn from even higher altitudes. These launch pattern shifts cause some more intact inactive payloads to accumulate below 500 km rather than above it, bringing them into natural compliance with the 25-year rule. The resulting increase in the lower shells due to natural decay also appears to drive commercial operators to return to their original launch behavior, reflecting the operators' distaste for both cost and collisions.

\subsection{Improved PMD compliance}

In this scenario compliance with the 25-year rule exogenously increases over time, reaching 100\% for all operator types in 2025 and remaining at 100\% thereafter. The scenario begins at the orbital environment state in 2018 and runs till 2030, with the ``other'' operator category repeating its 2018-2020 launch patterns. As in the lower-cost LEO access scenario, sectoral revenues and government budgets are projected forward at a 15\% annual growth rate and the cost of launch vehicles used by all operator types decreases at an annual rate of 1\% after 2020. 

\begin{figure}
\centering
\includegraphics[width=\linewidth]{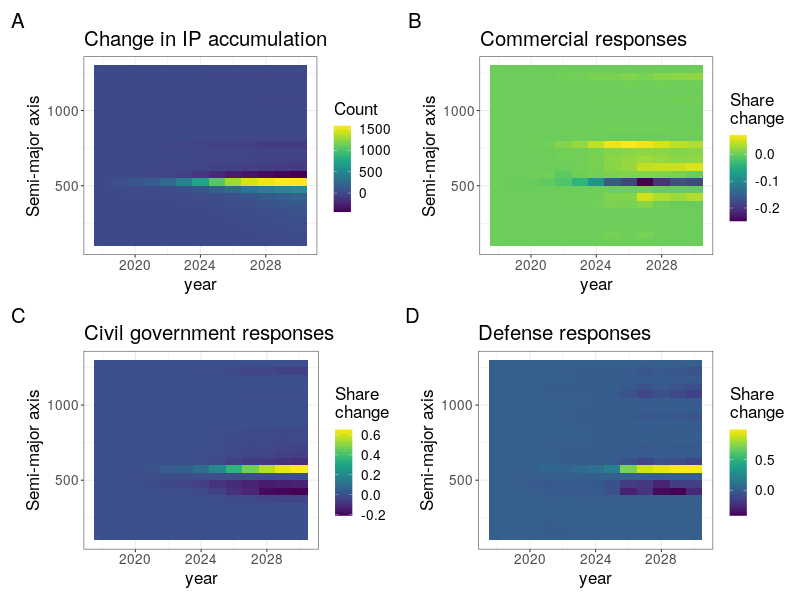}
\caption{Projected behavioral responses to increasing PMD compliance with the 25-year rule, reaching full compliance from 2025 and onwards.}  \label{fig:pmd_compliance}
\end{figure}

This scenario does not account for the cost of compliance to operators, and the model does not endogenize disposal decisions.\footnote{Prior studies such as \cite{brettle2019future} and \cite{brettle2020commercial} have examined the costs and benefits of increasing PMD compliance from an operator's perspective.} While this framework is not intended to produce long-run forecasts of the debris environment, it is worth considering these responses in light of analyses such as \cite{lewis2009fade} and \cite{liou2013leo}. Prior analyses have shown that greater PMD compliance with the 25-year rule will reduce debris accumulation substantially over the next 100-200 years by reducing the persistence of intact objects and any fragments they produce. Results from the integrated model indicate the presence of an additional behavioral channel. Moving intact inactive payloads to lower altitudes increases the expected collision rate at those altitudes. While the debris will decay relatively quickly, the expected increase in collisions will raise avoidance-related costs for operators and induce them to launch to other altitude shells. To the extent that nearby orbits are closer substitutes for each other and operators worry about decaying debris passing near their satellites, new launches will tend to shift to higher altitude shells. Such behaviors may have consequences for long-run debris accumulation patterns and the design of PMD guidelines.


\section{Limitations and future directions}
\label{limitsandfuture}

The validation results and scenarios considered point to several current limitations and directions for future work in this area. There are at least three categories of improvements to be made:

\begin{enumerate}
    \item better calibration of the PIB model to long-term debris environment simulations from higher-fidelity models;
    \item extending the second stage of the econometric model to incorporate more realistic substitution patterns;
    \item disaggregating time periods to produce larger samples for the first stage of the econometric model.
\end{enumerate}

The importance of calibrating the PIB model can be seen clearly in the orbital decay and intact payload accumulation patterns in the bottom row of Figure \ref{fig:model_validation4}. Calibrating the PIB model will also improve projections from the econometric model, as they depend on the state of the orbital environment.

A major limitation of the multinomial logit form in Equation \ref{CP_logit} is that it implies proportional substitution across orbital shells. That is, the ratio of choice probabilities for two shells, $j$ and $j'$, stays constant when an attribute of $j$ or $j'$ changes. This is only possible if both choice probabilities change by the same proportion. Such substitution may be realistic in some settings, but may also lead to unrealistic behavioral projections in others. This assumption can be relaxed through use of more flexible models, such as the probit or mixed logit \cite{trainbook}. Extending the second stage of the econometric model to a more flexible form can make projected substitution patterns more realistic at the cost of some additional model complexity. Mixed logit models may be especially promising as they can approximate a broader class of utility functions \cite{trainbook}. However, to the extent that operators are converging on or avoiding particular orbital shells due to common unmodeled factors, estimates of the parameters governing operators' responses to each other (i.e. $\beta_i$) may be unreliable. Examples of potential omitted factors relevant to all operators include radio spectrum availability, common orbit use cases, or direct market competition. Parameter identification in the presence of ambiguous strategic complementarity or substitution is challenging, particularly when multiple equilibria are possible \cite{timmins2007revealed}. Collecting more data on relevant orbit and operator characteristics along with developing approaches to identify these parameters will help improve second-stage model projections.

Estimates from the first stage of the econometric model suffer from a small sample size. Disaggregating the data to a finer timescale, e.g. monthly, would increase the sample size and may improve the precision of parameter estimates. This disaggregation may also introduce more dispersion in launch rates and require adjustment for zeros, either through statistical models which can account for overdispersion or an explicit model of the zeros. While it seems unlikely that monthly fluctuations in sectoral spending will drive large portions of launch behavior, the appropriate variables to include in these models at finer timescales is still unclear. 





\section{Conclusion}


This paper presents a framework for integrating theoretically-grounded behavioral models of launch decisions with multi-species physical models of the orbital environment. Such integrated models may reveal new features of proposed approaches to space traffic management and space sustainability. Proposals ranging from active debris removal and new PMD guidelines to orbital allocation and pricing schemes can be studied in a single model environment, and behavioral and mechanical effects can be quantitatively compared. Further work developing integrated models of orbit use will facilitate greater understanding of the interactions between satellite operators and the orbital environment.

\section*{Acknowledgments}

We are grateful to Gordon Lewis, Ethan Berner, and Teo Flesher for their assistance in locating and compiling data on satellite launch choice occasions and launch vehicle prices. We are also grateful to Becki Yukman and Space Foundation for providing us with data on sectoral revenues and government budgets.


\begin{thebibliography}{}

\bibitem{adilov2015economic}
Adilov N., Alexander P.J., Cunningham B.M., (2015). An economic analysis of earth orbit pollution. {\it Environmental and Resource Economics}, {\bf 60}(1), 81--98.

\bibitem{adilov2018kessler}
Adilov N., Alexander P.J. and Cunningham B.M., (2018). An economic ``Kessler Syndrome'': A dynamic model of earth orbit debris. {\it Economics Letters}, {\bf 166}, 79--82.

\bibitem{adilov2020model}
Adilov N., Alexander P.J. and Cunningham B.M., (2020). The economics of orbital debris generation, accumulation, mitigation, and remediation. {\it Journal of Space Safety Engineering}, {\bf 7}(3), 447--450.

\bibitem{bolduc1996doctor}
Bolduc D., Fortin B., Fournier M.A., (1996). The effect of incentive policies on the practice location of doctors: a multinomial probit analysis. {\it Journal of Labor Economics}, {\bf 14}(4), 703--732.

\bibitem{breiman2001forest}
Breiman L., (2001). Random forests. {\it Machine learning}, {\bf 45}(1), 5--32.

\bibitem{brettle2019future}
Brettle H., Forshaw J., Auburn J., Blackerby C. and Okada N., (2019). Towards a Future Debris Removal Service: Evolution of an ADR Business Model. In {\it 70th International Astronautical Congress}.

\bibitem{brettle2020commercial}
Brettle H., Ziadlourad A., Lindsay M., Forshaw J., Auburn J., (2020). Commercial Incentives for Debris Removal Services. In {\it 71st International Astronautical Congress}.

\bibitem{cameron1986models}
Cameron A.C., Trivedi P.K., (1986). Econometric models based on count data: comparisons and applications of some estimators and tests. {\it Journal of Applied Econometrics}, {\bf 1}(1), 29--53.

\bibitem{deaton1980aids}
Deaton A., Muellbauer J., (1980). An almost ideal demand system. {\it The American Economic Review}, {\bf 70}(3), 312--326.

\bibitem{DISCOS}
Database and information system characterising objects in space, (2021). \url{https://discosweb.esoc.esa.int/}. Accessed: 2021-03-19.

\bibitem{dobriban2018ridge}
Dobriban E., Wager S., (2018). High-dimensional asymptotics of prediction: Ridge regression and classification. {\it The Annals of Statistics}, {\bf 46}(1), 247--279.

\bibitem{envRep2020}
ESA Space Debris Office, (2020). ESA's Annual Space Environment Report. GEN-DB-LOG-00288-OPS-SD.

\bibitem{faa}
Federal Aviation Administration, (2021). Annual Compendium of Commercial Space Transportation. Available at: \url{https://www.faa.gov/space/additional_information/}

\bibitem{freeman2014measurement}
Freeman III A.M., Herriges J.A., Kling C.L., (2014). {\it The measurement of environmental and resource values: theory and methods}. Routledge.

\bibitem{hausman1984models}
Hausman J., Hall B.H., Griliches Z., (1984). Econometric models for count data with an application to the patents-R\&D relationship. {\it Econometrica}, {\bf 52}(4), 1219--1240.

\bibitem{hausman1995utility}
Hausman J.A., Leonard, G.K., McFadden, D., (1995). A utility-consistent, combined discrete choice and count data model assessing recreational use losses due to natural resource damage, {\it Journal of Public Economics}, {\bf 56}(1), 1--30.

\bibitem{hoerl1985ridge}
Hoerl A.E., Kennard R.W., Hoerl R.W., (1985). Practical use of ridge regression: A challenge met. {\it Journal of the Royal Statistical Society: Series C (Applied Statistics)}, {\bf 34}(2), 114-–120.

\bibitem{IADC2020}
Inter-Agency Space Debris Coordination Committee, (2020). Space Debris Mitigation Guidelines. {\it IADC-02-01}, Revision 2.


\bibitem{jonasetal}
Jonas A., Sinkevicius A., Flannery S., Swinburne B., Wellington P., Tsui T., Lalwani R., Faucette J.E., Nowak B., Shanker R., Pan K., Zlotnicka E., (2017). Space: Investment Implications of the Final Frontier. {\it Morgan Stanley Investment Report}.

\bibitem{jorgensen1988budgeting}
Jorgenson D.W., Slesnick D.T., Stoker T.M., (1988). Two-stage budgeting and exact aggregation. {\it Journal of Business \& Economic Statistics}, {\bf 6}(3):313--325.

\bibitem{krisko2011}
Krisko, P. H. (2011). Proper implementation of the 1998 NASA breakup model. {\it Orbital Debris Quarterly News}, {\bf 15}(4), 1-10.


\bibitem{lewis2009fade}
Lewis H.G., Swinerd G.G., Newland R.J., Saunders A., (2009). The fast debris evolution model, {\it Advances in Space Research}, {\bf 44}, 568--578.

\bibitem{liou2013leo}
Liou J.C., Anilkumar A.K., Bastida B., Hanada T., Krag H., Lewis H., Raj M., Rao M., Rossi A. and Sharma R., (2013). Stability of the future LEO environment–an IADC comparison study. In {\it Proceedings of the 6th European Conference on Space Debris} 723.

\bibitem{mansson2011poisson}
M\aa nsson K., Shukur G., (2011). A Poisson ridge regression estimator. {\it Economic modeling}, {\bf 28}(4), 1475--1481.

\bibitem{muller2017valuation}
Muller C., Rozanova O. and Urdanoz M., (2017). Economic valuation of debris removal. In {\it 68th International Astronautical Congress}.

\bibitem{nevo2011models}
Nevo A., (2011). Empirical models of consumer behavior. {\it Annual Review of Economics}, {\bf 3}(1), 51--75.

\bibitem{pantanowitz2009missing}
Pantanowitz A., Tshilidzi M., (2009). Missing Data Imputation Through the Use of the Random Forest Algorithm. {\it Advances in Computational Intelligence}, 53--62. Springer Berlin Heidelberg.

\bibitem{raoetal2020}
Rao A., Burgess M.G., Kaffine D., (2020). Orbital-use fees could more than quadruple the value of the space industry, {\it Proceedings of the National Academy of Sciences}, {\bf 117}(23), 12756--12762.

\bibitem{rouillon2020economic}
Rouillon S., (2020). A Physico-Economic Model of Low Earth Orbit Management. {\it Environmental and Resource Economics}, {\bf 77}(4), 695--723.

\bibitem{somma2017statistical}
Somma G. L., Colombo C., Lewis H.G., (2017). A statistical LEO model to investigate adaptable debris control strategies, {\it 7th European Conference on Space Debris, ESA/ESOC}, ESA Space Debris Office.

\bibitem{sommathesis}
Somma G.L., (2019). Adaptive remediation of the space debris environment using feedback control. {\it University of Southampton, Doctoral Thesis}, 166pp.

\bibitem{spacereport}
Space Foundation, (2021). The Space Report. Available at: \url{https://www.thespacereport.org}

\bibitem{timmins2007revealed}
Timmins C., Murdock J., (2007). A revealed preference approach to the measurement of congestion in travel cost models, {\it Journal of Environmental Economics and Management}, {\bf 53}(2), 230--249.

\bibitem{train1998recreation}
Train K. E., (1998). Recreation demand models with taste differences over people. {\it Land Economics}, 230-239.

\bibitem{trainbook}
Train K. E., (2009). {\it Discrete choice methods with simulation}. Cambridge University Press.



\end{thebibliography}
\end{document}